\documentstyle[11pt,moriond,epsfig]{article}

\def\be{\begin{equation}}
\def\eq{\end{equation}}
\def\bea{\begin{eqnarray}}
\def\eea{\end{eqnarray}}

\def\ff{\mathrm{f\bar f}}
\def\ee{\mathrm{e^{+}e^{-}}}
\def\ll{\mathrm{l^{+}l^{-}}}
\def\eeff{\ee\rightarrow\ff}
\def\tautau{\tau^{+}\tau^{-}}
\def\mumu{\mu^{+}\mu^{-}}
\def\tautau{\tau^{+}\tau^{-}}
\def\tt{\tau^{+}\tau^{-}}
\def\mm{\mu^{+}\mu^{-}}
\def\qq{\mathrm{q}\overline{\mathrm{q}}}

\def\bb{{\mathrm{b}\overline{\mathrm{b}}}}
\def\cc{{\mathrm{c}\overline{\mathrm{c}}}}
\def\eeqq{\ee\rightarrow\qq}

\def\eell{\ee\rightarrow\ll}
\def\eeee{\ee\rightarrow\ee}
\def\spr{s'}
\def\Zprime{Z'}
\def\MZplim{m_{\Zprime}^{lim}}
\def\GeV{\mathrm{GeV}}
\begin{document}
\vspace*{4cm}
\title{SEARCH FOR  EXOTIC  PHYSICS WITH FOUR-FERMION COUPLING  AT LEP}

\author{ SABINE RIEMANN }
\address{DESY Zeuthen, Platanenallee 6, \\ D-15738 Zeuthen, Germay}

\maketitle\abstracts{
Preliminary results of measurements of fermion-pair production at LEP2
are used to derive limits on new physics phenomena.
Combinations of the cross sections  and asymmetries 
of the 4 LEP collaborations
are interpreted in terms four-fermion contact interactions, exchange of Z$'$ 
and leptoquarks. Results on the search 
for extra dimensions are also presented.
} 

\section{Introduction: Fermion-Pair Production at LEP2}
In the years from 1995 to 2000 LEP delivered 
an integrated luminosity of
700~pb$^{-1}$
to each experiment
at c.m.s. 
energies between 130 GeV and 209 GeV. This allows
a precise measurement of the process $\eeff$ 
and therefore a test of the
Standard Model
at high energies.

\begin{figure}[htbp]
 \begin{center}
   \begin{tabular}{lr}
     \includegraphics[width=0.45\textwidth,height=0.6\textwidth]{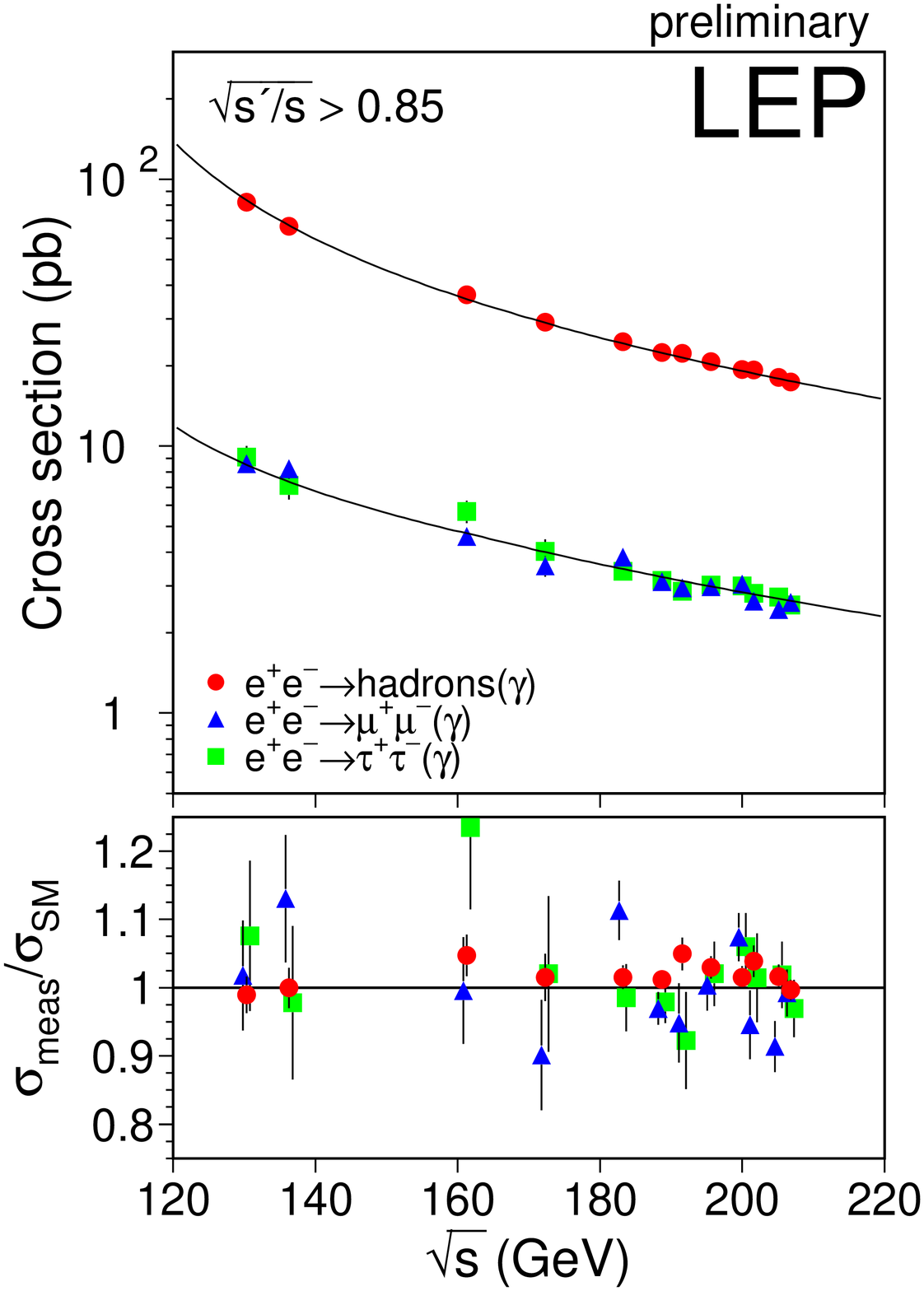}&
     \includegraphics[width=0.45\textwidth,height=0.6\textwidth]{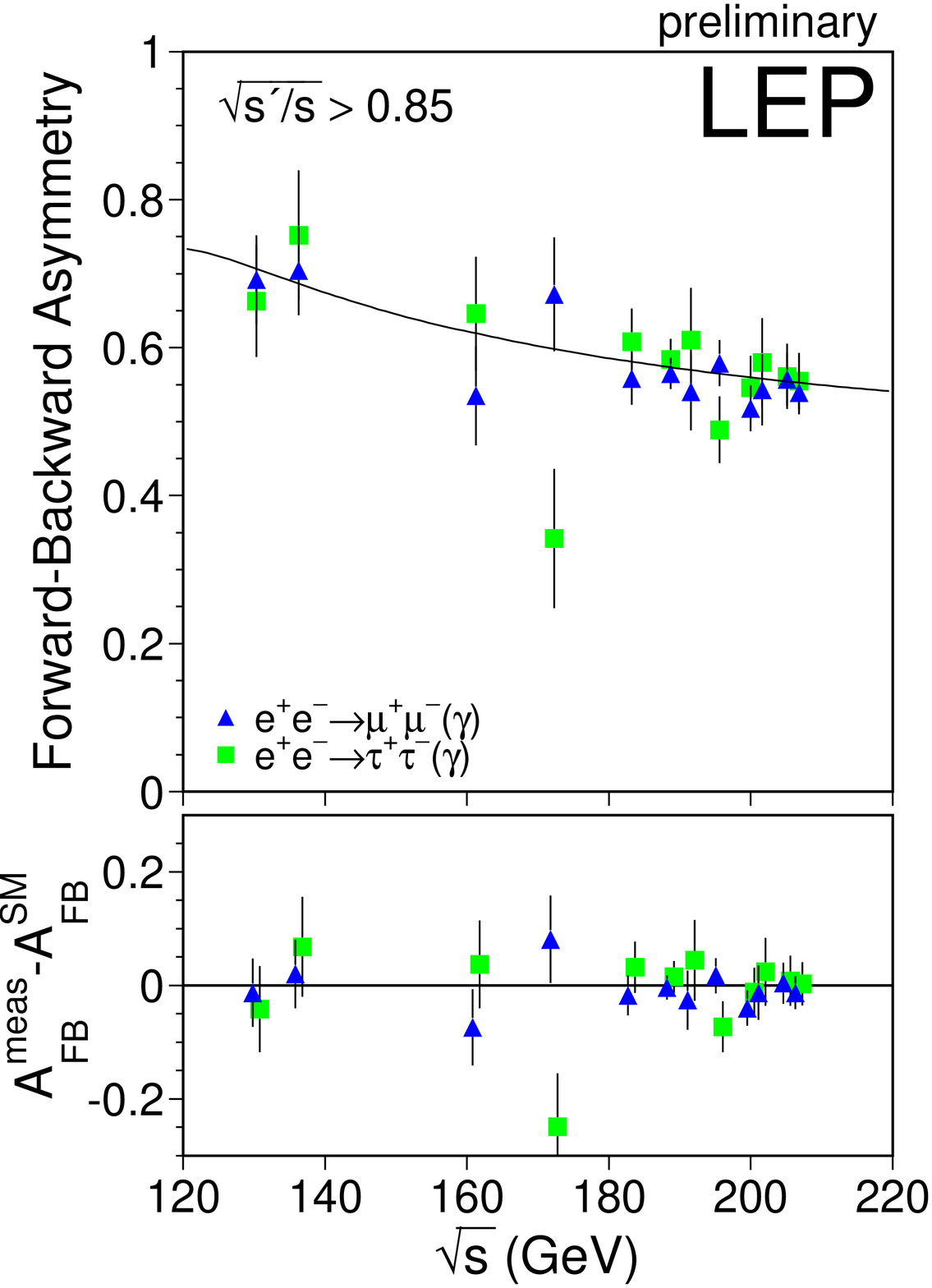}
   \end{tabular}
   \caption{Combined LEP results on cross sections and
   forward-backward asymmetries for $\qq$, $\mumu$  and $\tautau$ final states
   as a function of centre-of-mass energy.}
 \label{fig:xs-afb}
 \end{center}
\end{figure}

\begin{figure}[htbp]
 \begin{center}
   \begin{tabular}{lr}
     \includegraphics[width=0.45\textwidth]{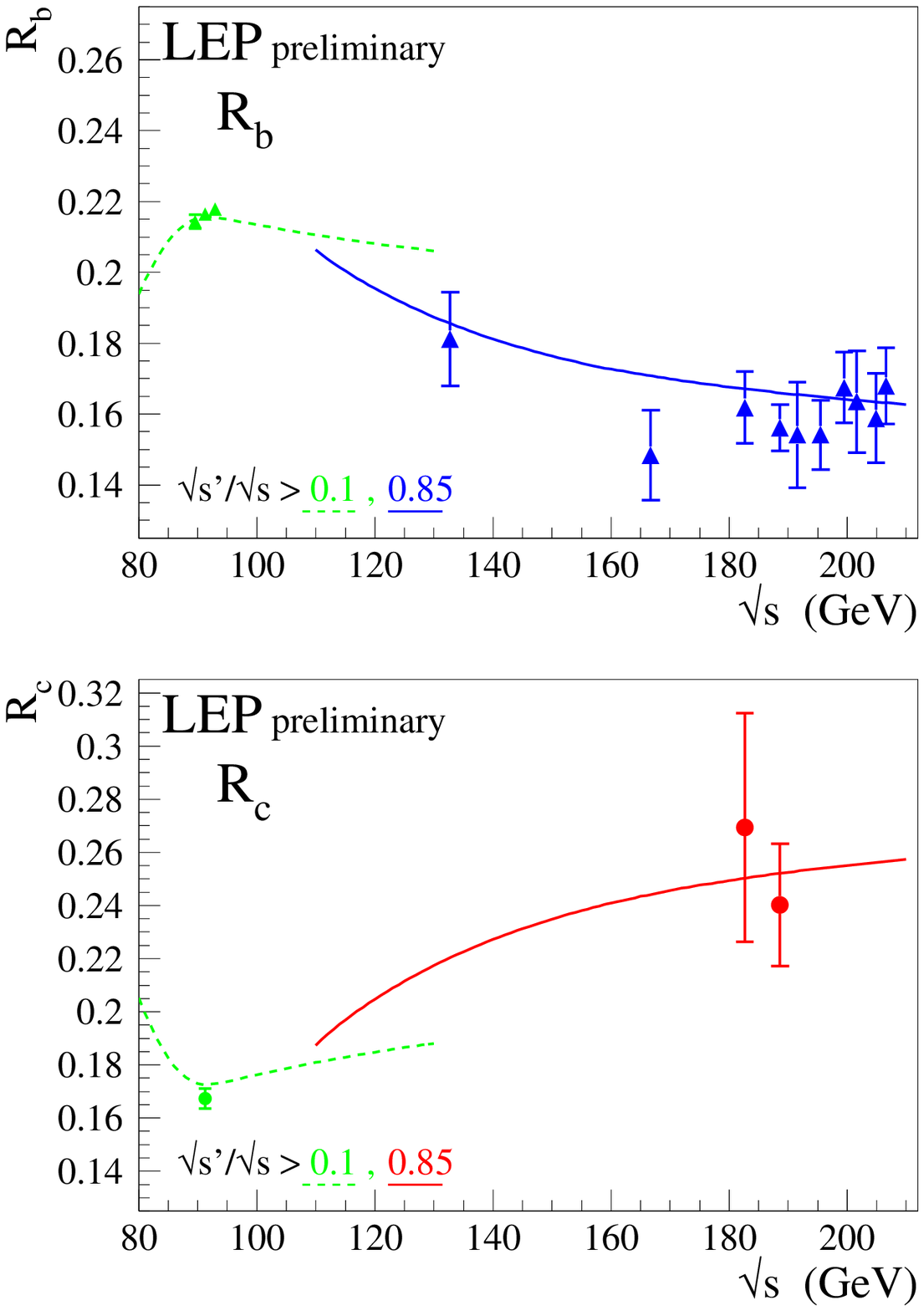}&
     \includegraphics[width=0.45\textwidth]{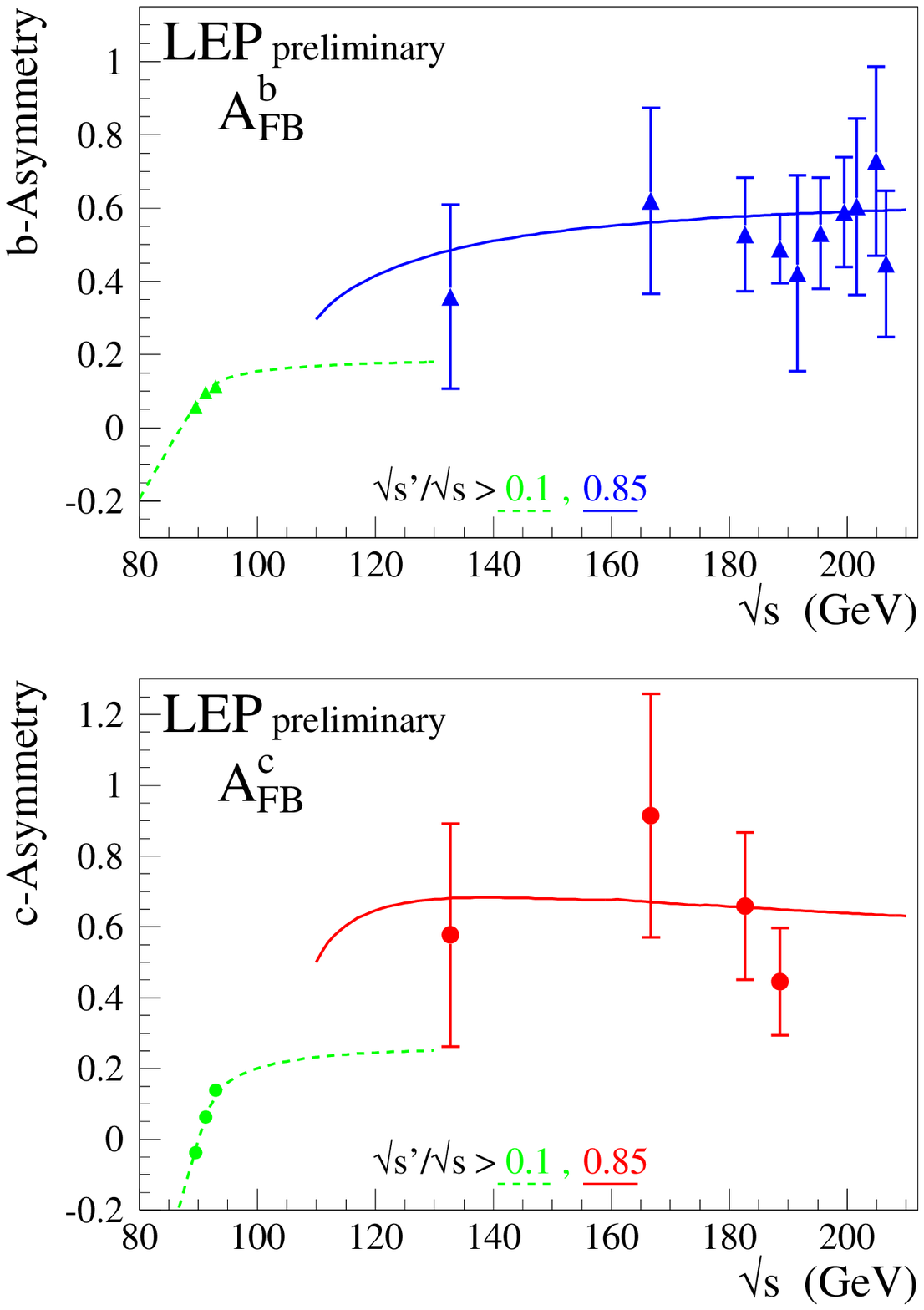}
   \end{tabular}
   \caption{Combined LEP results on $R_b$, $R_c$ and on
   forward-backward asymmetry for $\bb$  and $\cc$ final states
   as a function of centre-of-mass energy.}
 \label{fig:hv}
 \end{center}
\end{figure}

The individual experiments' analyses of cross-sections
are combined for the $\eeqq,~\mm$, $\tt,~\bb$, $~cc$
channels, forward-backward asymmetry measurements are combined for
the $\mumu$, $\tautau$, $\bb,~\cc$ final states.
The averages are made for the
samples of events with $\sqrt{\spr}>0.85\sqrt{s}$
in order to suppress a 'pollution' of potential signals of
new physics   by radiative returns to the Z peak.
A detailed description of the combination 
procedure can be found in Reference 1 
and references
therein.
The         results                     are shown in
Figure~\ref{fig:xs-afb}. The combination of published and 
preliminary results of the $R_q = \sigma_q/\sigma_{had}$ and $A_{FB}^q$
measurements~\cite{ref:lep-comb} 
are shown in   
Figure~\ref{fig:hv}.
\section{Interpretations}
Although  
all measurements confirm  the Standard Model
predictions so far, new physics
phenomena could be hidden in
small deviations from  theoretical expectations.
A fit to the combined data
including a variety of new physic models may reveal
new effects  or to place limits on their parameters.
All limits presented here are given at the 95\% confidence level.
\subsection{Contact Interactions}
\begin{figure}[htbp]
  \begin{tabular}{lcr}
   \includegraphics[width=0.32\textwidth,height=0.45\textwidth]{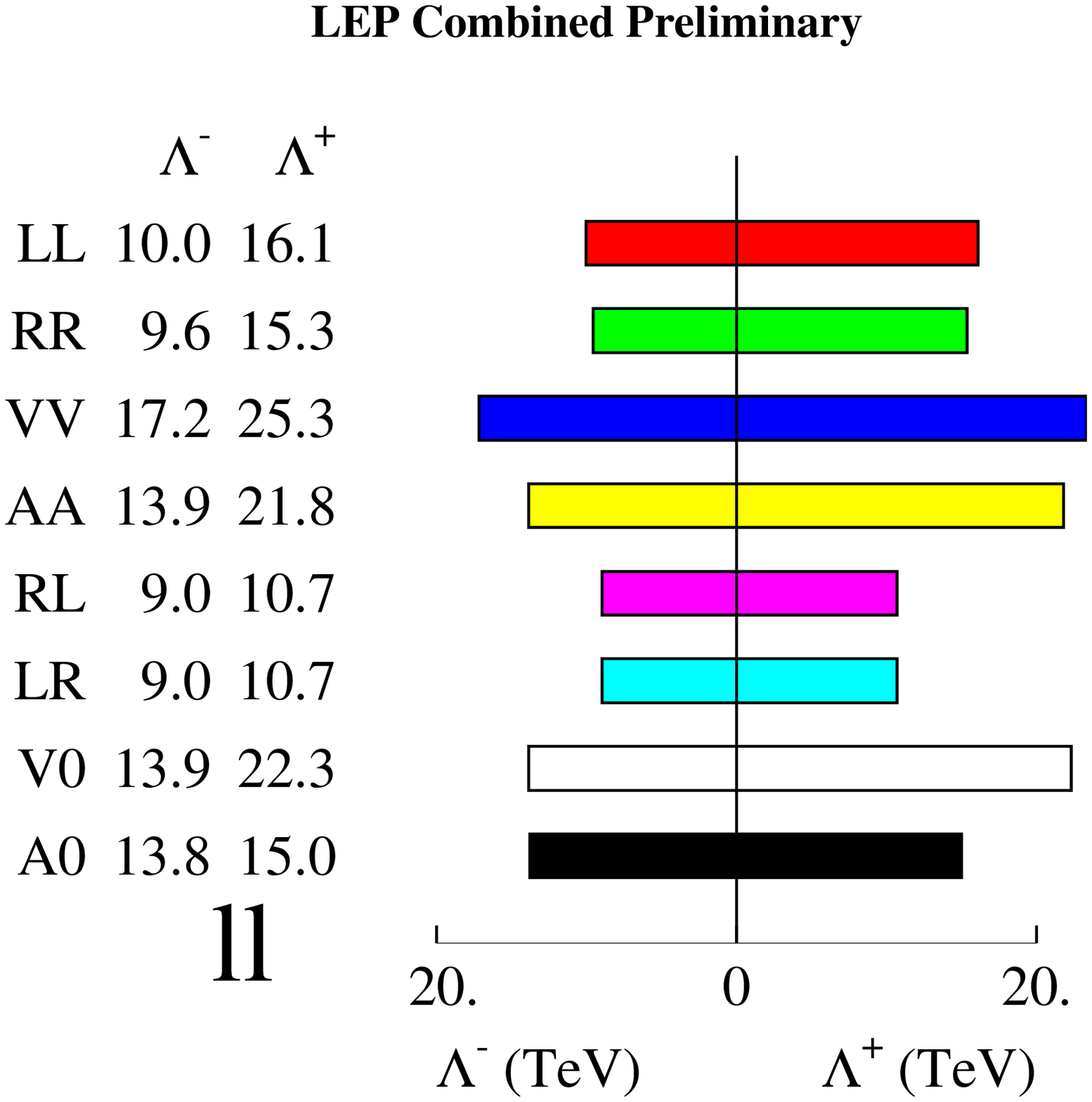}&
   \includegraphics[width=0.32\textwidth,height=0.45\textwidth]{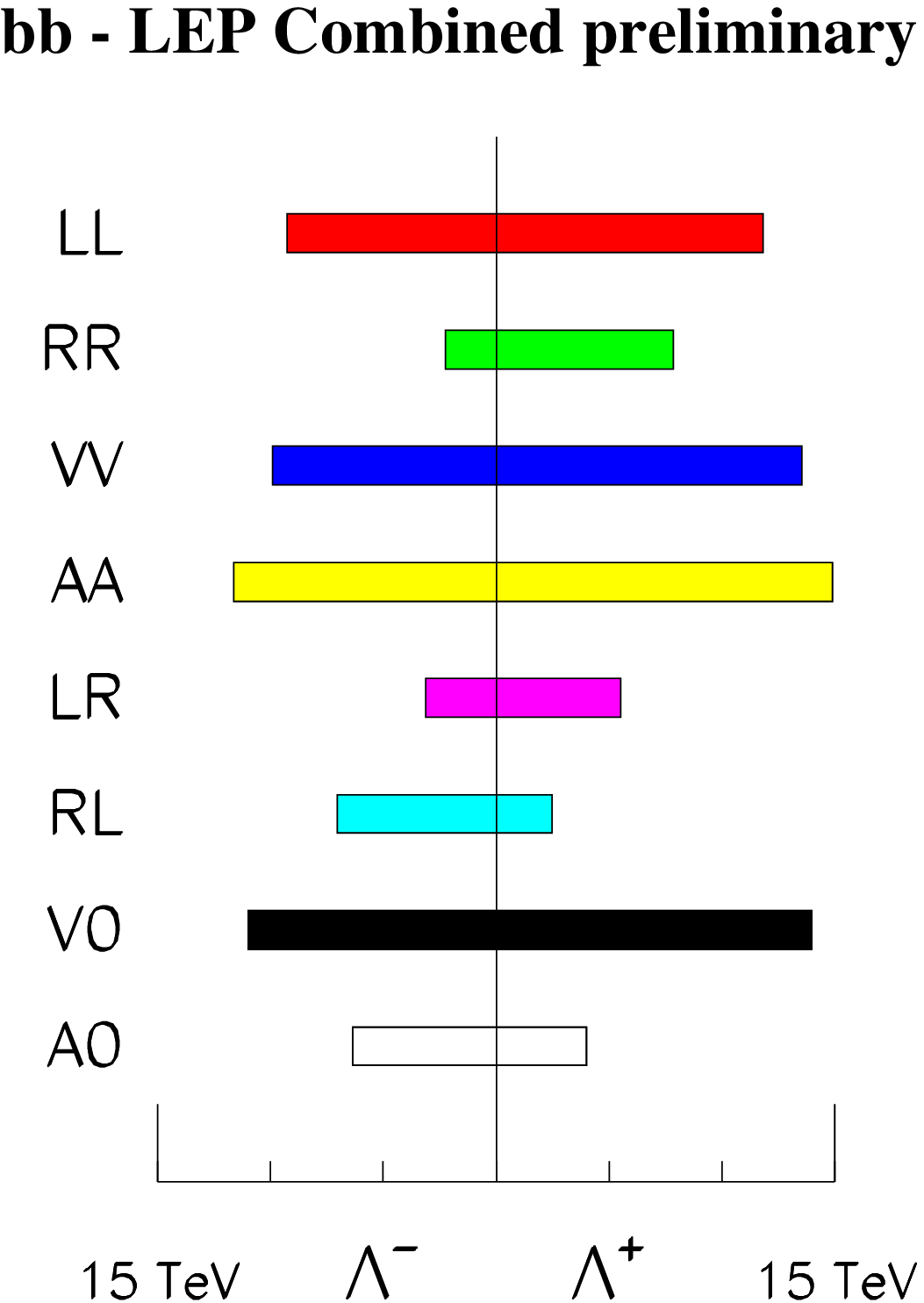}&
   \includegraphics[width=0.32\textwidth,height=0.45\textwidth]{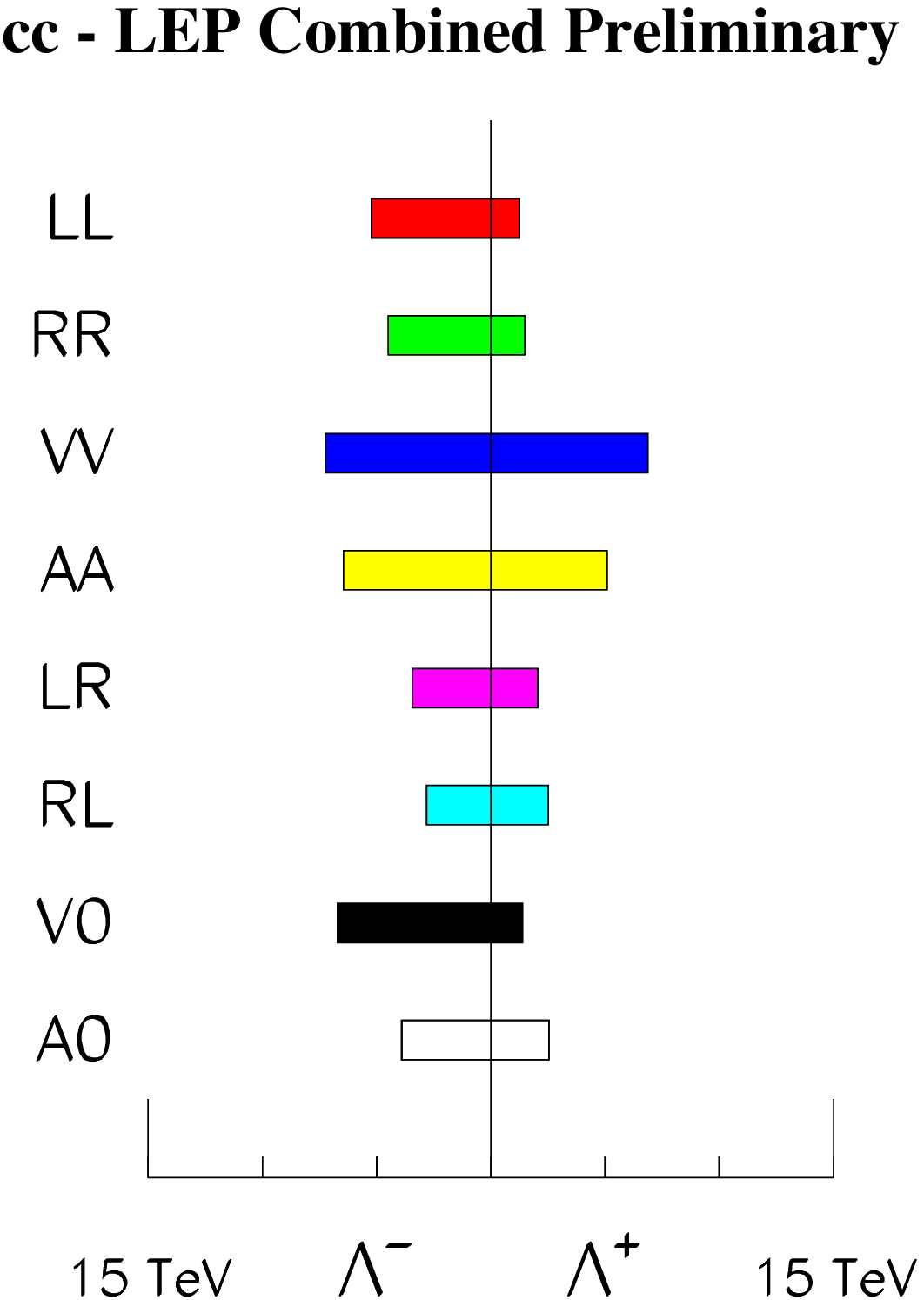}
  \end{tabular}
\caption{
Limits on the scale of Contact Interactions in $\eell,~\bb$
         and $\cc$ using  LEP combined results
         from 133 to 209$\GeV$. }
 \label{fig:cibbcc}
\end{figure}

Four fermion contact interactions~\cite{ref:ci-thry}
parametrise interactions beyond the 
Standard Model by an effective scale, $\Lambda$.
 Models can be defined by assuming
different helicity couplings between initial and final
state currents with either constructive (+) or destructive(-)
interfence between Standard Model and contact term.
No deviations $\epsilon \propto 1/\Lambda^2$
from the Standard Model predictions have been obtained for the considered
final states.
Bounds on $\epsilon$ were derived and converted into
limits on $\Lambda^+$ and $\Lambda^-$. The results are shown in
Figures~\ref{fig:cibbcc} for averaged and combined $\mumu$ and $\tautau$
final states assuming lepton universality, for $\bb$ and for $\cc$ final
states.
\subsection{Extra Z Bosons}
Additional heavy gauge bosons, Z$'$, are predicted by many theories.
The combined hadronic and leptonic cross sections  and leptonic
forward-backward asymmetries were used to fit the data to models from an
E$_6$ GUT or from left-right symmetries~\cite{ref:zp-thry}.
No evidence for the existence of a Z$'$ was found.
Table~\ref{tab:zp} presents 
limits on the mass
$m_{\mathrm{Z'}}$ for the special models $\chi$, $\psi$, $\eta$ and  L-R
and for the
Sequential Standard Model (SSM), that proposes a Z$'$
with exactly the same couplings to fermions like the Z.
\begin{table}[t]
\caption{95\% confidence level lower limits on the $\Zprime$ mass and
$\chi$, $\psi$, $\eta$, L-R and SSM models.}\label{tab:zprime_mass_lim}
\vspace{0.4cm}
\begin{center}\renewcommand{\arraystretch}{1.2}
\begin{tabular}{|cc|c|c|c|c|c|}\hline
 \multicolumn{2}{|c|}{Model}          & $\chi$  & $\psi$ & $\eta$ & L-R  & SSM  \\ \hline \hline
$\MZplim$           & ($\GeV/c^{2}$) & 715     & 478    & 454    &  
862 & 2090 \\ \hline \end{tabular}
\end{center} \label{tab:zp}
\end{table}
\subsection{Leptoquarks}\label{subsec:lq}
Leptoquarks mediate lepton-quark transitions. Following the basics in
References 4, 
scalar (S) and vector (V) leptoquarks
could be exchanged in the u- or t-channel of the process $\eeqq$.
Assuming a coupling of electromagnetic strength, 
$g=\sqrt{4\pi\alpha_{em}}$, and one type of leptoquarks much lighter than
the others,
limits on the masses of leptoquarks are 
derived and presented in Table~\ref{tab:lq-mass}.
In most cases the kinematical limits of searches for direct leptoquark
production at LEP
are exceeded by this indirect search. Further, the bounds presented here,
complement the searches at HERA~\cite{ref:lq-hera}.
\begin{table}[tp]
 \caption{\it{Lower limits on the LQ mass assuming
           $g_{L,R}=\sqrt{4\pi \alpha}$. 
           For $ \tilde{S}_{1/2}$(L) no limit can be set.}}
   \label{tab:lq-mass}
  \renewcommand{\arraystretch}{1.12}
  \begin{center}
   \begin{tabular}{|l|r||l|r|}
     \hline
        LQ type  & $m_{\mathrm{LQ}}^{min}$($\GeV/c^2$)
      & LQ type  & $m_{\mathrm{LQ}}^{min}$($\GeV/c^2$) \\
      \hline
      \hline
      $S_0(\mathrm{L})\rightarrow \mathrm{e u}$                      &  789 &
      $V_{1/2}(\mathrm{L})\rightarrow \mathrm{e d}$                  &  305\\
      $S_0(\mathrm{R})\rightarrow \mathrm{e u}$                      &  639 &
      $V_{1/2}(\mathrm{R})\rightarrow \mathrm{e u,~ e d}$            &  227 \\
      $\tilde{S}_0(\mathrm{R})\rightarrow \mathrm{e d}$              &  210 &
      $\tilde{V}_{1/2} (\mathrm{L}) \rightarrow \mathrm{e u}$        &  176 \\
      $S_1(\mathrm{L})\rightarrow \mathrm{e u,~ ed}$                 &  364 &
      $V_0(\mathrm{L})\rightarrow \mathrm{e \bar{d}}$                & 1070 \\
      $S_{1/2}(\mathrm{L})\rightarrow \mathrm{e \bar{u}}$            &  189 &
      $V_0(\mathrm{R})\rightarrow \mathrm{e \bar{d}}$                &  167 \\
      $S_{1/2}(\mathrm{R})\rightarrow \mathrm{e \bar{u},~e \bar{d}}$ &  240 &
      $\tilde{V}_0(\mathrm{R})\rightarrow \mathrm{e \bar{u}}$        &  497 \\
      $\tilde{S}_{1/2}(\mathrm{L})\rightarrow \mathrm{e \bar{d}}$    & --   &
      $V_1(\mathrm{L})\rightarrow \mathrm{e \bar{u},~e \bar{d}} $    &  664 \\
     \hline     
  \end{tabular}
\end{center}
\end{table}
\subsection{Extra Dimensions}
Recently, theories of quantum gravity in extra spatial 
dimensions~\cite{ref:ed-thry}
have been an elegant means to  escape
the hierarchy problem: 
Gravitons are considered to propagate in a compactified higher dimensional
space while Standard Model particles exist in the usual 3+1 space-time 
dimensions. The Planck mass in the 'normal' 4 dimensions is  related to
a fundamental scale $M_D \approx \Lambda_{ew}$ in n+4 dimensions by
$M_{Pl}^2 = R^n M_D^{n+2}$,
where $R$ is the compactification radius of extra dimensions.
The exchange of       gravitons could contribute to
the processes
$\eeff,~ \gamma \gamma,~\mathrm{ZZ},~\mathrm{W}^+\mathrm{W}^-$.
The search for indirect effects of low scale gravity was performed by
all LEP collaborations seperately~\cite{ref:ed-lep} and has not yet 
been combined.
The best sensitivity can be
reached in  the process
$\eeee$.  Including this channel, preliminary results
on $M_S\propto M_D$ range between 0.98~TeV/c$^2$ and 1.17~TeV/c$^2$
for a   theory parameter $\lambda=-1$ and between
0.84~TeV/c$^2$ and 1.06~TeV/c$^2$ for $\lambda=+1$.
     \section*{Acknowledgments}
I thank my colleagues from the LEP collaborations and the LEPEWWG
Fermion Pair Subgroup.

\end{document}